\documentclass[conference]{IEEEtran}
\IEEEoverridecommandlockouts
\usepackage{cite}
\usepackage{amsmath,amssymb,amsfonts}
\usepackage{algorithmic}
\usepackage{graphicx}
\usepackage{float}
\usepackage{textcomp}
\usepackage{xcolor}
\usepackage{caption}
\usepackage{booktabs}
\usepackage{url}
\usepackage{hhline}
\usepackage{array}
\def\BibTeX{{\rm B\kern-.05em{\sc i\kern-.025em b}\kern-.08em
    T\kern-.1667em\lower.7ex\hbox{E}\kern-.125emX}}
\begin{document}

\title{Classifying Issues in Open-source GitHub Repositories\\}

\author{
    \IEEEauthorblockN{Amir Hossain Raaj\IEEEauthorrefmark{1}, Fairuz Nawer Meem\IEEEauthorrefmark{2}, Sadia Afrin Mim\IEEEauthorrefmark{2}}   \IEEEauthorblockA{\IEEEauthorrefmark{1,}\IEEEauthorrefmark{2,} \IEEEauthorrefmark{3}Department of Computer Science, George Mason University}
    \IEEEauthorblockA{
    \IEEEauthorrefmark{1}araj20@gmu.edu,
    \IEEEauthorrefmark{2}fmeem@gmu.edu,
    \IEEEauthorrefmark{3}safrinmi@gmu.edu,
    }
    }

\maketitle

\begin{abstract}
GitHub is the most widely used platform for software maintenance in the open-source community. Developers report issues on GitHub from time to time while facing difficulties. Having labels on those issues can help developers easily address those issues with prior knowledge of labels. However, most of the GitHub repositories do not maintain regular labeling for the issues. The goal of this work is to classify issues in the open-source community using ML \& DNN models. There are thousands of open-source repositories on GitHub. Some of the repositories label their issues properly whereas some of them do not. When issues are pre-labeled, the problem-solving process and the immediate assignment of corresponding personnel are facilitated for the team, thereby expediting the development process. In this work, we conducted an analysis of prominent GitHub open-source repositories. We classified the issues in some common labels which are: API, Documentation, Enhancement, Question, Easy, Help-wanted, Dependency, CI, Waiting for OP’s response, Test, Bug, etc. Our study shows shat DNN models outperforms ML models in terms classifying GitHub issues showing upto 83.75\% accuracy.
\end{abstract}

\section{Introduction}
One essential tool for organizing tasks, improvements, and bugs in software projects is GitHub issues. An efficient method of categorizing and naming these problems is essential to preserving a well-organized workflow. Teams may more efficiently prioritize and arrange work by classifying issues with labels, such as bugs, feature requests, or documentation updates, which will result in a more efficient process.\\
Labels improve team communication on a project as well. They are able to promptly inform all contributors of the state of each problem, letting them know if it is pending input, is being worked on, or requires assistance. This transparency lowers the possibility of task duplication or oversight and promotes improved teamwork among members. Furthermore, having the ability to effectively filter concerns becomes essential in large projects with a large number of difficulties. By designating problems with phrases like "critical" for serious difficulties or "low priority" for less urgent ones, labels help in this process. By focusing contributors and maintainers on the most important tasks first, this prioritization increases overall productivity.\\
Another important benefit of efficient labeling is automated issue triage. Tools that attach labels to new issues automatically based on keywords can streamline the initial handling process and save time. Large projects, where manual triage might become a bottleneck, can benefit especially from this. Project teams' ability to analyze data is also improved by labeling. Maintainers can learn more about prevalent problem types, track how much time is spent on each category, and pinpoint possible areas for improvement by classifying issues using labels. Planning next development and improving project management techniques both benefit greatly from this data.\\
In the context of open-source projects, labeling can be extremely important to community involvement. By identifying tasks that are appropriate for novices, labels such as "good first issue" can draw in new contributors. This promotes the development of the community surrounding the project as well as aids in the integration of new participants. Accurate issue labeling in GitHub is crucial for reasons more than just organization. It facilitates automatic triage, boosts productivity through improved prioritization, increases communication, offers insightful data via analytics, and promotes community involvement. A more effective and prosperous project management environment is facilitated by each of these elements.\\
In this work, we categorized unlabelled issues from the 100 most-starred repositories from GitHub. We collected only issues with the 9 default GitHub labels which are: Documentation, Enhancement, Question, Help Wanted, Duplicate, Invalid, Bug, Good First Issue, Won't Fix. We collected around 568,000 issues with those labels both opened and closed from those repositories. The main contributions of this work are:
\begin{itemize}
    \item Creating a huge dataset of issues with default GitHub labels.
    \item Categorizing unlabeled issues using various Machine Learning \& Deep Learning algorithms
\end{itemize}

\section{Related works}

There have been some similar or related work in this sector. Previously, Williams et al. proposed a method to use the source code change history for improving bug finding techniques \cite{williams2005automatic}. This technique repeatedly checks through the change of the code that implies the type of fixed issues that are commonly solved in repositories. The novelty of the work is they automatically mined the repositories and showed the rankings. They showed warnings and likely bugs for Apache web servers and Wine. However, their work did not explicitly classify the bugs analyzing natural languages used for the bug descriptions and their choice of tooling is different.\\ \\
Ahmed et al. \cite{ahmed2014predicting} presented predictions of bug categories based on analysis of software repositories. This work is very closely similar to ours. They used K Nearest Neighbor and Naïve Bayes for predicting types of bugs. They classified issues in categories like standard related, function related, user interface related, logic related etc. Their work showed best accuracy with Naïve Bayes classifier. However, their work contains a relatively smaller set of categories and does not work with fairness repositories.\\ \\
Later, Kumaresh et al. \cite{kumaresh2015mining} categorized defects mining software repositories. They also analyzed online bug repositories like JIRA, Bugzilla, Eclipse etc. using text clustering for classifying data and showed F-measure, Accuracy etc. Their classification of bugs according to text clustering might help project managers to assign correct bugs to the correct developers.\\ \\
Recently, Hamdy et al. \cite{hamdy2022deep} performed a deep mining of open-source repositories. They mined from large scale bug-tracking systems Jira, Bugzilla etc. Five powerful deep learning architectures, including CNNs, LSTMs, GRUs, and hybrid CNN-LSTM/GRU models were trained for the purpose of categorizing bug reports according to severity ( for example - high, medium, low). They showed how five of these algorithms can be used to show higher accuracy in different contexts. However, their work shows classification based on severity and mostly focuses on bugs present in large scale bug tracking systems, not about open-source repositories.\\ \\
Nagwani et al. \cite{nagwani2012clubas} proposed an algorithm and Java based tool for software bug classification using bug attributes’ similarities. The classification mainly works by text clustering, frequent term calculations and taxonomic terms mapping techniques. Algorithms used for clustering are Naïve Bayes, Naïve Bayes Multinomial, J48, Support Vector Machine and Weka’s classification using clustering algorithms. Similarly to other work, this work does not dedicatedly analyze the landscape of the issues present in open-source ethical repositories.
\section{Methodology}
Here in this section, we will describe how we accomplished this work step by step. The details are described in subsequent subsections.
\subsection{Data Collection}
\begin{figure}
\centering
  \includegraphics[width=8cm,height=7cm]{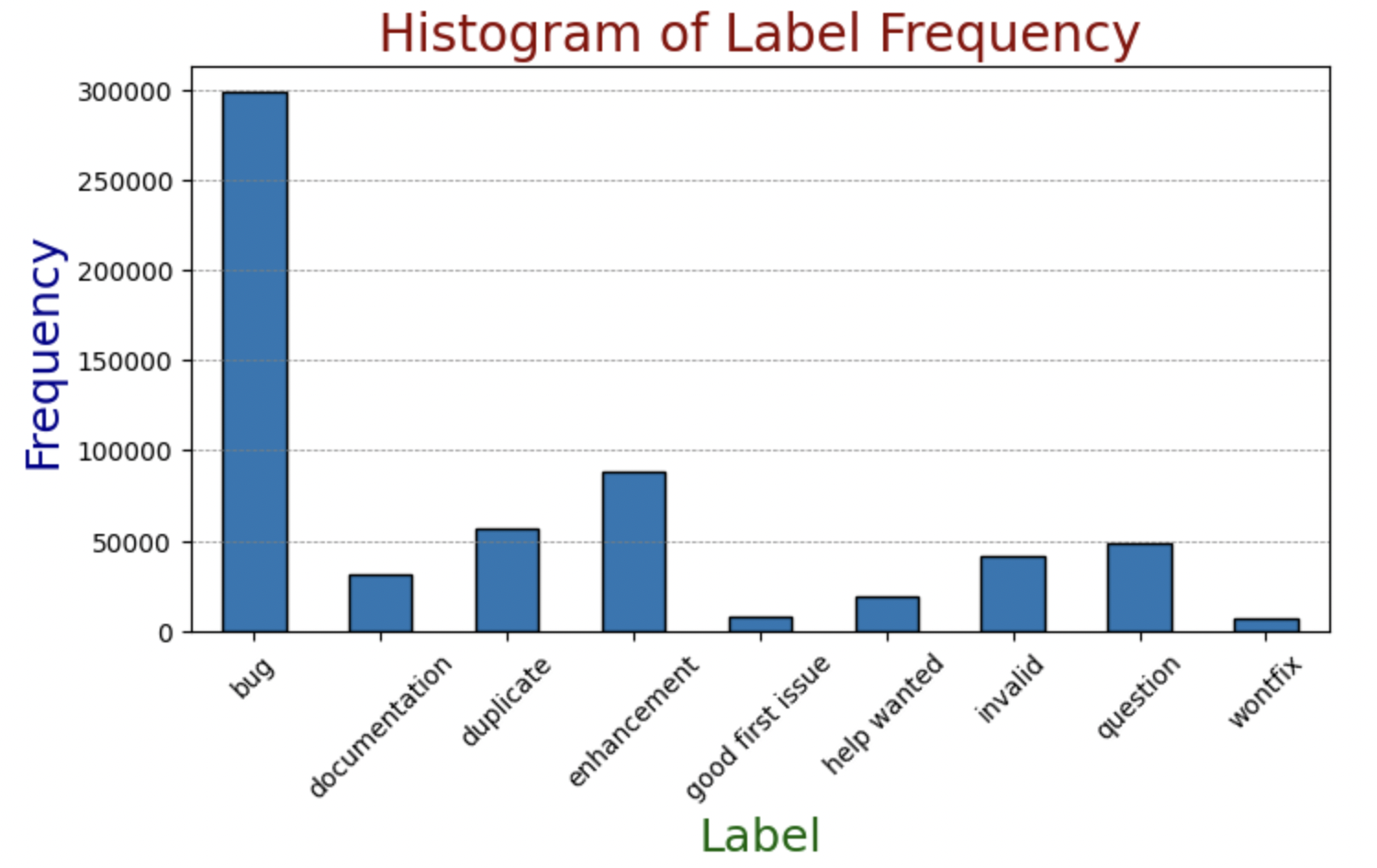}
  \caption{Frequency of labels}
  \label{hist}
\end{figure}

First we selected the top 100 most-starred GitHub repositories for this work. We considered both open and closed issues. We considered only the default 9 GitHub issue labels which are: which are: Documentation, Enhancement, Question, Help Wanted, Duplicate, Invalid, Bug, Good First Issue, and Won't Fix. We aimed to create a single-class classification for them. The frequency of labels in data are shown in Figure~\ref{hist}.

\subsection{Data Preprocessing}
\subsubsection{Setup and Initialization}
In order to preprocess our data, we imported the essential modules for dealing with the operating system, object serialization, and regular expressions. We imported several libraries for specific tasks: \textit{numpy} and \textit{pandas} for data manipulation, \textit{spacy} and \textit{nltk} for natural language processing, \textit{matplotlib} for data visualization, \textit{BeautifulSoup} for HTML parsing, and \textit{sklearn} for machine learning.\\
To access our input files, we established a connection between the Colab environment and Google Drive, allowing us to immediately access files stored on Google Drive. Next, we established a collection of file paths from which data would be retrieved and indicated the location where the merged data will be stored.\\
We created an empty list to contain dataframes. Subsequently, we iterated through each file path, extracted the contents into a DataFrame, and added it to the list. Afterwards, we merged all DataFrames into a unified DataFrame and saved it as a CSV file.\\
We have implemented the \textit{nltk} library for the purpose of natural language processing, as well as the \textit{lxml} library for parsing XML and HTML using the Python programming language. We optimized the performance of the SpaCy English language model by disabling specific components and initializing a stemmer. After that, we established functions to preprocess textual data, encompassing the tasks of purging HTML material, eliminating undesirable characters, and standardizing the text.\\
\subsubsection{Data Preparation}
Initially, we imported the merged CSV file into a DataFrame and eliminated rows where the 'Body' column contains missing values (NaN). Next, we utilized the preprocess\_text function on designated columns to remove any inconsistencies and normalize the text data. We designated a new destination file directory for the cleaned data and saved the DataFrame to a CSV file.\\
This is how our script was designed to preprocess and merge CSV files that include our dataset. It focuses on cleaning and standardizing text fields in order to facilitate subsequent analysis.
\section{Dataset}
This is the final dataset \footnote{\url{https://drive.google.com/file/d/16ltnYYOPP9qJ6CiPzYTDJ3KxHOM_daqK/view?usp=sharing}} that we used for our experiments.
\section{Classification Models}
A classification model is a machine learning system that predicts categorical class labels from input data. These models employ advanced algorithms to examine the correlation between input features and predetermined categories, with the goal of precisely classifying new data points into one of these categories. Classification models are widely used in diverse fields, including spam detection, picture recognition, and medical diagnosis. We used several classification models for our work.
\subsection{ML Based}
\subsubsection{Naive Bayes}
Naive Bayes is a classification model that uses Bayes' Theorem and assumes that the predictors are independent of each other. The algorithm computes the likelihood of each category based on a given set of input characteristics and chooses the category with the highest likelihood as the output. Naive Bayes is highly efficient and successful when applied to huge datasets, making it a popular choice for text classification tasks.
\subsection{Random Forest}
The Random Forest algorithm is a type of ensemble learning approach that builds several decision trees during the training process. It then determines the final predicted class by selecting the class that appears most frequently among the predictions made by the individual trees. The model enhances classification accuracy by employing bagging and introducing feature randomness during the construction of each tree, thus ensuring robustness against overfitting. Random Forest is highly effective in addressing a diverse array of issues, particularly ones that entail intricate interactions and structures within the data.
\subsection{Deep Learning Based}

\subsubsection{RNN}
Recurrent Neural Networks (RNN) are a type of neural networks specifically created to handle sequential input. They are able to capture dynamic temporal patterns by utilizing a hidden state that functions as a type of memory. Recurrent Neural Networks (RNNs) are highly beneficial for applications such as speech recognition and natural language processing. However, they frequently encounter difficulties in handling long-range relationships, mostly because of the vanishing gradient problem.
\subsubsection{LSTM}
Long Short-Term Memory (LSTM) networks are a type of Recurrent Neural Networks (RNNs) that are specifically designed to effectively deal with long-range dependencies in data. LSTMs incorporate gate mechanisms that control the information flow, enabling the network to selectively retain or discard information. LSTMs are particularly efficient for intricate sequential tasks, such as predicting time series and generating text.
\subsubsection{GRU}
Gated Recurrent Units (GRU) are a modified version of Long Short-Term Memory (LSTM) units that aim to streamline the model while achieving comparable performance. GRUs integrate the forget and input gates into a unified "update gate" and combine the cell state and hidden state, leading to a more efficient learning process. GRUs are employed in comparable situations to LSTMs, frequently demonstrating comparable performance with lower computational burden.
\subsection{Transformers}
\subsubsection{Bert}
BERT, developed by Google, is an innovative model in the field of natural language processing. The model employs the Transformer architecture and incorporates a novel approach of bidirectional training, enabling it to simultaneously examine the complete context of a sentence from both left to right and right to left. BERT is highly proficient in tasks such as question answering, language inference, and sentiment analysis.
\subsubsection{Roberta}
RoBERTa improves upon BERT by adjusting critical hyperparameters in the model and conducting training with a significantly larger dataset for extended durations. The modifications entail the elimination of the purpose of predicting the following sentence, conducting training with larger mini-batches, and use longer sequences. RoBERTa has demonstrated superior performance compared to BERT on multiple NLP benchmarks.
\subsubsection{Distilbert}
DistilBERT is an optimized iteration of BERT that prioritizes speed and efficiency, making it more suited for tasks that have limited computational capabilities. It maintains 95\% of BERT's efficacy in language comprehension tasks while achieving a 60\% increase in speed. DistilBERT is developed by a technique known as distillation, where a smaller model is trained to replicate the functionality of a bigger, pre-trained model.
\section{Experimental Setup \& Results}
The results of our project are segmented into three distinct subsections based on the modeling approaches: ML-based, Deep Learning-based, and Transformer-based models, each tailored to evaluate their efficacy in addressing the project's objectives. We have summarized the performance of various machine learning models in Table 1

\begin{table}[ht]
\centering
\caption{Model Performance Comparison}
\label{tab:model_accuracy}
\begin{tabular}{@{}lc@{}}
\toprule
\textbf{Model}                      & \textbf{Accuracy (\%)} \\ \midrule
Naive Bayes (using title)           & 59                     \\
Naive Bayes (using body)            & 55                     \\
Random Forest (using title)         & 71                     \\
Random Forest (using body)          & 72                     \\
RNN                                 & 73.35                  \\
LSTM                                & 79.18                  \\
GRU                                 & 83.75                  \\
BERT                                & 78                     \\
ROBERTA                             & 78                     \\ \bottomrule
\end{tabular}
\end{table}
\subsection{ML Based}
In configuring our machine learning models, we adopted a single label selection approach to simplify multi-label datasets into more manageable single-label instances. Text data was processed using TF-IDF vectorization, which enhances the representation of important textual features by reducing the impact of common words across documents. We split the dataset into training and testing sets with an 80:20 ratio, ensuring substantial data for training while maintaining enough for effective model validation. For the Random Forest classifier, we chose to utilize 100 trees, balancing the ability to learn complex patterns with the need to avoid overfitting, thus optimizing both the learning process and the model’s generalization to new data.
\subsubsection{Naive Bayes}
The bar chart presents the performance metrics of the Naive Bayes classification model, specifically Precision, Recall, and F-1 Score, for various issue labels (such as bug, documentation, duplicate, etc.) in our dataset.\\
Precision, indicated by the blue bars, is the proportion of accurately predicted positive observations out of the total anticipated positives. High precision is characterized by a minimal occurrence of false positives.
Recall, denoted by the orange bars, is the proportion of accurately predicted positive observations out of all observations in the actual class. High recall is associated with a low incidence of false negatives.
The F-1 Score, depicted by the gray bars, is a metric that calculates the weighted average of Precision and Recall. The F-1 Score achieves its optimal value of 1 when both precision and recall are flawless, and it reaches its lowest value of 0 when both precision and memory are poor.
\begin{figure}[H]
    \centering
    \includegraphics[width=8cm]{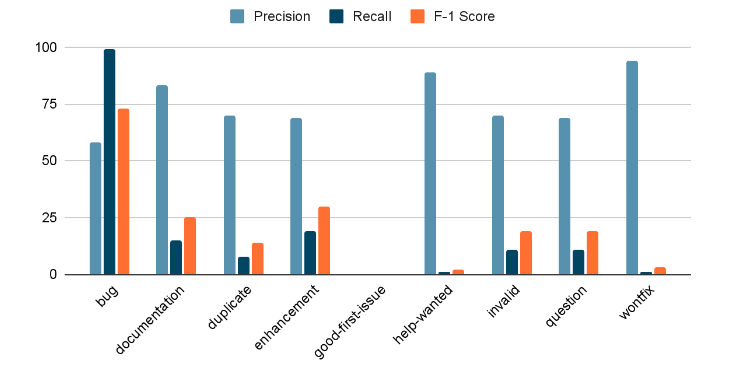}
    \caption{Classification report of Naive Bayes Classifier}
    \label{NaiveBayes}
\end{figure}
We can see that the label "bug" exhibits the highest level of precision, while its recall and F-1 Score are at a moderate level. The label "documentation" exhibits a high level of recall, but its precision and F-1 Score are comparatively lower. The labels "good first issue" and "wontfix" exhibit a very low level of precision and recall, leading to correspondingly low F-1 Scores. The "question" label has consistently poor scores across all three criteria, suggesting that the model faces difficulties in this particular label.
\subsubsection{Random Forest}
The bar chart presents the performance metrics of the Random Forest classification model.
\begin{figure}[H]
    \centering
    \includegraphics[width=8cm]{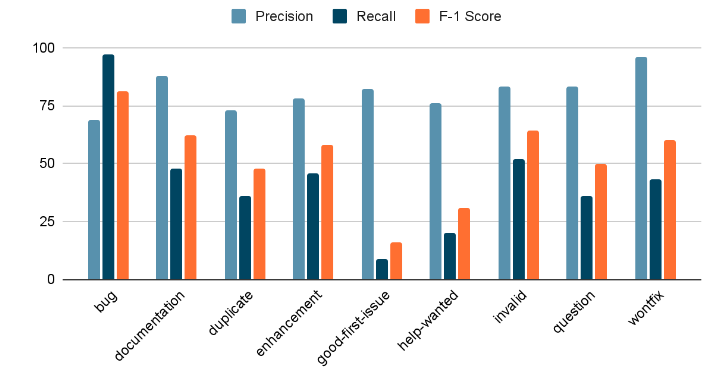}
    \caption{Classification report of Random Forest Classifier}
    \label{RandomForest}
\end{figure}
Here, the labels "bug" and "documentation" have a high level of Precision but a lower level of Recall, indicating that the model is careful yet precise in its predictions for these labels. The label "enhancement" demonstrates a well-balanced performance with commendable precision and recall, as seen by the F-1 score. The labels of "Good first issue" and "help wanted" exhibit poor ratings across all metrics, indicating that they present significant challenges for the model. The Precision of the "Question" label is low, meaning that the model frequently misclassifies other categories as "question." However, the Recall is high, suggesting that the model correctly detects a large proportion of actual "question" instances. The "Wontfix" label has moderate levels of Precision and Recall, with a somewhat lower F-1 Score. This indicates a balanced performance, albeit at a lower level compared to certain other issue labels.

\subsection{Deep Learning Based}
In our deep learning experiments, we configured several hyperparameters to optimize model performance. We limited the maximum vocabulary size to 10,000 to balance the trade-off between computational complexity and model accuracy. The dataset was divided into a training and testing split of 70:30, ensuring sufficient data for learning while retaining enough to validate the model's generalization capability. We trained our models for 50 epochs, allowing the models enough iterations to learn from the data effectively. Each training batch contained 64 samples, optimizing the balance between memory usage and model updating frequency.

The architecture of our neural networks included hidden layers with a dimensionality of 100, which provided ample capacity to capture complex patterns in the data without excessively increasing computational demands. We employed CrossEntropyLoss for the loss function, adjusted by class weights to handle imbalances in the dataset, ensuring fair attention to all classes during training. 

\begin{figure}[H]
    \centering
    \includegraphics[width=8cm]{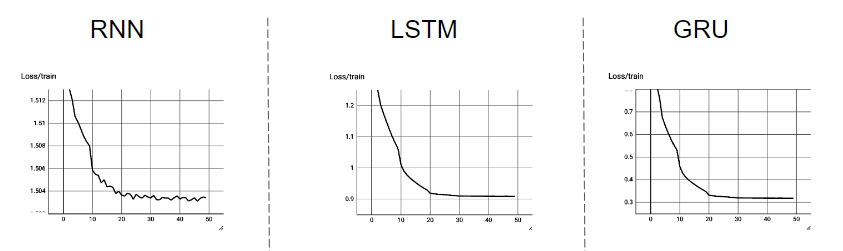}
    \caption{Loss per epoch of Deep Learning models without Class Weights Adjustment}
    \label{withoutWeight}
\end{figure}

\begin{figure}[H]
    \centering
    \includegraphics[width=8cm]{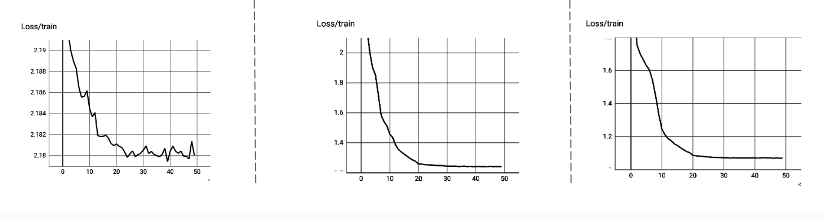}
    \caption{Loss per epoch of Deep Learning models with Class Weights Adjustment}
    \label{withWeight}
\end{figure}

The learning rate was set at 0.001, with a scheduler implementing a step decay every 10 epochs, where the learning rate is multiplied by a factor of 0.1 (gamma). This strategy helps in fine-tuning the model by reducing the learning rate over time, preventing overfitting on the training data and aiding in convergence to a more stable model state.
The images here display a comparison of training loss curves for three different deep learning models: RNN (Recurrent Neural Network), LSTM (Long Short-Term Memory), and GRU (Gated Recurrent Unit).\\
Fig \ref{withoutWeight} represents training loss curves without class weight adjustment.
The recurrent neural network (RNN) exhibits a progressive decrease in loss across 50 epochs, indicating the occurrence of learning. However, there is some instability in the loss, as evidenced by occasional tiny upward fluctuations. The LSTM model demonstrates a consistent and gradual decrease in loss, demonstrating a continuous and effective learning process throughout the epochs. Additionally, it achieves a smaller loss value compared to RNN, indicating superior performance.
The GRU model has the most quick learning with a smooth and sharp decrease, resulting in the lowest loss value. This indicates that the GRU model has the most effective learning process in this context.\\
On the other hand, Fig \ref{withWeight} shows training loss curves with class weight adjustment, likely to address class imbalance.
The RNN has a more unpredictable loss curve when class weights are adjusted, displaying greater variations in training loss. This suggests that the model may struggle to effectively fit the data when the weights are modified. The LSTM model demonstrates a consistent reduction in loss even when adjusting for class weights, however the initial and final loss values are larger compared to the unweighted case. GRU, similar to LSTM, exhibits a steady and smooth learning trajectory by adapting its weights. Furthermore, it demonstrates a robust ability to reduce loss even when there are variations in class weights.\\ 
In general, the graphic indicates that GRU and LSTM, due to their intricate structures that may capture long-term relationships, are achieving better results compared to the simpler RNN model, regardless of whether class weight changes are used or not. The loss curves provide an indication of the learning progress of each model in predicting the data during training. A model's performance improves as the curve becomes smoother and lower. The class weight adjustments are likely incorporated to enhance the performance of the model on datasets with imbalanced classes. It appears that LSTM and GRU are better able to handle these adjustments compared to RNN.
\subsection{Transformer Models}
The images show two line charts representing the performance of two Transformer models, BERT and RoBERTa, over the course of three epochs. Each chart tracks three metrics: Training Loss, Validation Loss, and Accuracy.
\subsubsection{BERT}
\begin{figure}[H]
    \centering
    \includegraphics[width=8cm]{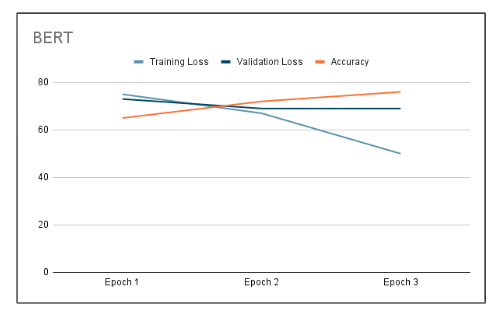}
    \caption{Losses and accuracy over epochs of BERT}
    \label{bert}
\end{figure}
The training Loss for BERT exhibits an initial high value which gradually reduces across epochs, suggesting that the model is effectively acquiring knowledge from the training data. The validation loss (shown by the orange line) first drops, but later reaches a plateau or slightly increases. This observation may indicate the onset of overfitting or that the model has already extracted all the information it can from the data. The accuracy of the model, represented by the red line, improves over time, indicating that the predictions made by the model are becoming more accurate as the training process advances.
\subsubsection{Losses and accuracy over epochs of RoBERTa}
\begin{figure}[H]
    \centering
    \includegraphics[width=8cm]{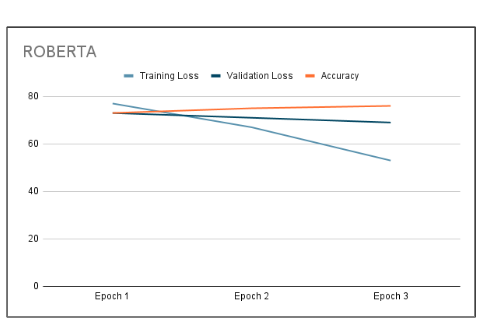}
    \caption{Losses and accuracy over epochs of RoBERTa}
    \label{roberta}
\end{figure}
In the case of RoBERTa, the training loss exhibits a consistent decline, similar to BERT, indicating enhanced ability to learn from the training data. The validation loss exhibits a decrease, mirroring the behavior of the training loss, indicating that the model is effectively generalizing. The accuracy is roughly constant and does not exhibit a substantial increase. The observed values are generally greater than those of BERT, indicating a potential superiority in accurate predictions. However, this discrepancy could also be attributed to variations in the y-axis scales.

\section{Conclusion}
In conclusion, this project successfully demonstrated the potential of automated labeling of GitHub issues using machine learning and deep learning techniques, particularly in the context of open-source software repositories. We meticulously categorized issues into commonly encountered labels like API, Documentation, Enhancement, and others, addressing a significant challenge in many GitHub repositories which lack systematic labeling. The classification models employed, ranging from ML-based to advanced Transformer-based models, not only optimized the issue handling process but also enhanced the assignment and management of tasks within project teams. Through this work, we have laid a groundwork that not only accelerates developmental workflows but also facilitates more effective project management and contributor involvement in the open-source community. This system stands to benefit a wide array of repositories by providing a scalable solution to the perennial problem of unstructured issue management.

\bibliographystyle{ieeetr}
\bibliography{bibliography}
\end{document}